\documentstyle[12pt]{article}
\newcommand\be{\begin{equation}}
\newcommand\ee{\end{equation}}
\newcommand\bea{\begin{eqnarray}}
\newcommand\eea{\end{eqnarray}}
\newcommand\ket[1]{|#1\rangle}
\newcommand\bra[1]{\langle #1|}

\newcommand{\fatalpha}{{\bf \alpha \kern -0.44em \alpha}}
\newcommand{\fatsigma}{{\bf \sigma \kern -0.54em \sigma}}
\newcommand{\tpchi}{{\bf \chi \kern -0.35em \chi}}
\newcommand{\llambda}{{\bf \lambda \kern -0.45em \lambda}}

\bibliography{plain}
\title{\bf  Investigation of quantum roulette}\vspace{20mm}
\author{ S. Salimi
  \thanks{Corresponding author: E-mail:shsalimi@uok.ac.ir} , \
  M.M. Soltanzadeh
  \thanks{E-mail:msoltanzadeh@uok.ac.ir}
 \\ {\small Department of Physics,
University of Kurdistan, Sanandaj 51664, Iran.} }  \pagebreak

\vspace{20mm}
\begin{document}
\maketitle \vspace{15mm}
\begin{abstract}
In this paper, by using permutation matrices as a representation of
symmetric group $S_N$ and Fourier matrix, we investigate quantum
roulette with an arbitrary $N$-state. This strategy, which we
introduce, is general method that allows us to solve quantum game
for an arbitrary $N$-state. We consider the interaction between the
system and its environment and study the effect of the depolarizing
channel on this strategy. Finally, as an example we employ this
strategy for quantum roulette with $N=3$.


{\bf PACs Index: 03.65.Ud }
\end{abstract}

\vspace{70mm}
\newpage
\section{Introduction}
Recent research in quantum computation, communication and
cryptography has focussed on the physical aspect of information. In
the most of the cases quantum description of the system provides
advantages over the classical situation. For example, Simon's
quantum algorithm  \cite{Simon} to identify the period of a function
chosen by a oracle is more efficient than any deterministic or
probabilistic algorithm, Shor's polynomial time quantum algorithm
\cite{Shor} for factoring and the quantum protocols for key
distribution devised by Bennet and Brassard \cite{Benet} and Ekert
\cite{Ekert} are qualitatively more secure against eavesdropping
than any classical cryptographic system.

Game theory is the study of decision making in conflict situation.
Recently, quantum game theory \cite{3,4,5,6,7,8,9,10,11,12,13, Ren}
has been investigated, which discusses versions of some classical
game \cite{14} where new rules that make explicit use of quantum
mechanics lead to new solutions. D.A. Meyer \cite{3} demonstrated
that in a classical two-person zero-sum strategic game, if one
person adopts a quantum strategy, then he has a better chance of
winning the game. And based on these work, Xiang-Bin Wang, L.C. Kwek
et al. \cite{15} extended this case by replacing the coin which has
only two possible states (namely head and tail) with a roulette with
$N=2^m$ ($m=1,2,...$) states, and concluded that quantum strategies
can also be more successful than classical ones; Jing-Ling Chen,
L.C. Kwek and C.H. Oh \cite{16} studied noisy quantum game. In this
paper we investigate quantum roulette with arbitrary  $N$ states. To
solve this problem we use permutation matrices as a representation
of symmetric group $S_N$ and Fourier matrix. This strategy is
general and one can generalize easily to any integer $N$ which for
example we employ it for quantum roulette with $N=3$. Also, in the
end we consider the interaction between the system and its
environment and investigate depolarizing channel on this strategy.

The organization of the paper is as follows. In Section 2, we give a
brief outline of a quantum coin-tossing game . In Section $3$, we
generalize a quantum coin-tossing by replacing the coin, which has
only two possible state, with a roulette with  arbitrary $N$ state
and we employ this strategy for a quantum game with $N=3$, also in
the end of this section  we investigate the effect of decoherence
channel, depolarizing channel,  on this strategy. The paper is ended
with a brief conclusion.

\section{A quantum coin-flipping game}
In this section we review Meyer's strategies to play with a single
coin. The classical coin has only two possible states, head and
tail. It is natural to define a two-dimensional Hilbert space $H_2$
with basis $\ket{H}$ and $\ket{T}$ (the symbols $H$ and $T$ denote
head and tail, respectively) which could be represented by:
\begin{equation}
\ket{H}=\left(
             \begin{array}{c}
               1 \\
               0 \\
             \end{array}
           \right),\qquad \ket{T}=\left(
             \begin{array}{c}
               0 \\
               1 \\
             \end{array}
           \right).
\end{equation}
The player strategies represent by matrices of $2\times 2$
\begin{equation}
N=\left(\begin{array}{cc}
      1 & 0 \\
      0 & 1 \\
    \end{array}
  \right), \qquad F=\left(
    \begin{array}{cc}
      0 & 1 \\
      1 & 0 \\
    \end{array}
  \right)
\end{equation}
correspond to not flipping and flipping the coin, respectively.
Owing to $N$ and $F$, we can construct a density matrix $D$ as
follows:
\begin{equation}
D=\frac{1}{2}(N+F)=\frac{1}{2}\left(
                                \begin{array}{cc}
                                  1 & 1 \\
                                  1 & 1 \\
                                \end{array}
                              \right), \qquad \mbox{Trace}(D)=1.
\end{equation}
It is easy to verify that $D$ commutes with $N$ and $F$, i.e.,
\begin{equation}
[N,D]=[N,F]=0.
\end{equation}
With due attention to the unitary of  $N$ and $F$, from above
equation we would have an identity
\begin{equation}\label{e1}
D=(1-p)NDN^{\dagger}+pFDF^{\dagger},
\end{equation}
which is independent upon the parameter  $p\in [0,1]$ which is
probability that the player flips the coin.

The general pure state if a quantum coin is
\begin{equation}
\ket{\psi}=\cos(\frac{\theta}{2})\ket{H}+e^{i\phi}\sin(\frac{\theta}{2})
\ket{T}
\end{equation}
where corresponding density matrix is given by
\begin{equation}
\rho=\ket{\psi}\bra{\psi}=\frac{1}{2}\left(
                                       \begin{array}{cc}
                                         1+\cos(\theta) & e^{-i\phi}\sin(\theta) \\
                                         e^{i\phi}\sin(\theta) & 1-\cos(\theta) \\
                                       \end{array}
                                     \right).
\end{equation}
Now Alice and Bob come to play a coin-tossing game,  such that
Alice utilize a classical probabilistic strategy in which she
flips the coin with probability $p$, but  Bob could control this
game by quantum strategies. The game accomplish in four step as follows:\\
\textbf{Step1:} Let us the initial state of the coin which is placed
by Alice be $\ket{\psi_0}$, thus its density matrix is given by
\begin{equation}
\rho_0=\ket{\psi_0}\bra{\psi_0}.
\end{equation}
\textbf{Step2:} Bob acts on coin by a quantum strategy (unitary
transformation $U_1$), then the state of the coin becomes

\begin{equation}
\rho_1=U_1\rho_0U_1^{\dagger}=D,
\end{equation}
where
$$U_1= (\ket{\lambda_0},
\ket{\lambda_1})=\frac{1}{\sqrt{2}}\left(
                                \begin{array}{cc}
                                  1 & 1 \\
                                  1 & -1 \\
                                \end{array}
                              \right),\ \mbox{if}\
                              \rho_0=\ket{H}\bra{H}
$$
\begin{equation}
U_1= (\ket{\lambda_1}, \ket{\lambda_0})=\frac{1}{\sqrt{2}}\left(
                                \begin{array}{cc}
                                  1 & 1 \\
                                  -1 & 1 \\
                                \end{array}
                              \right),\ \mbox{if}\
                              \rho_0=\ket{T}\bra{T}.
\end{equation}
In the above equation the $\ket{\lambda_0}$ and $\ket{\lambda_1}$
are eigenvectors of density matrix $D$ with eigenvalues
$\lambda_0=1$ and $\lambda_1=0$, respectively.

\textbf{Step3:} Alice continues to play with a classical strategy,
namely, Alice employs a convex sum of unitary (deterministic)
transformation, i.e., he either flips the coin using the
transformation $F$ with probability $p$ or lets the coin rest in its
original state (using the identity transformation $N$) with
probability $(1-p)$. Thus by using (\ref{e1}), one can know  that
Alice's classical strategy does not change the density matrix of the
coin, i.e.,
\begin{equation}
\rho_2=(1-p)N\rho_1N^{\dagger}+pF\rho_1F^{\dagger}=\rho_1=D.
\end{equation}
\textbf{Step4:}  Finally, Bob could control the game by an
appropriate unitary transformation, $U_2=U_1^\dagger$, i.e., because
the density matrix is still $D$ he can adopt an unitary matrix $U_2$
to transform it into the  state which he wants to.

\section{Investigation of quantum roulette with arbitrary  $N$ state}
In this section we generalize a quantum coin-tossing by replacing
the coin, which has only two possible state (namely head and tail),
with a roulette with $N$ state.

 Here we consider Hilbert space $H$ as a quantum roulette with
 $N$-dimensional
and we indicate $N$ basis of quantum roulette with $\ket{k}$ for
$k=1,2,...,N$, where could be represented by the following matrices:
\begin{equation}
\ket{1}=\left(
\begin{array}{c}
 1 \\
 0 \\
 \vdots \\
0 \\
0 \\
\end{array}
\right),\quad \ket{2}=\left(
\begin{array}{c}
0 \\
1 \\
\vdots \\
0 \\
0 \\
\end{array}
\right), \quad \cdots \quad\ket{N}=\left(
                                \begin{array}{c}
                                  0 \\
                                  0 \\
                                  \vdots \\
                                  0 \\
                                  1 \\
                                \end{array}
                              \right).
\end{equation}
Thus, the roulette has  $N$ state which Alice has a choice  $N!$
possible flips corresponding to all the possible permutation of the
state $\{1,2,3,...,N\}$ to itself, where is called symmetric group
$S_{N}$. Therefore the explicit matrix form of operators are
permutation matrices as  a representation of symmetric group
$S_{N}$. By using Ref.\cite{sagan}, if $\pi\in S_{N}$ then the
permutation matrix $X(\pi)=(x_{i,j})_{N\times N}$ is defined as
follows:
\[
x_{i,j} = \left\{
\begin{array}{ll}
1 & \mbox{if $ \pi(j)=i$}\\
0 & \mbox{otherwise.}
\end{array}
\right.
\]
that it contains only zeros and ones, with unique one in every row
and column. Now we let $X^i$, for $i=0,1,2,...N!-1$,  as permutation
operators, then we can construct the density matrix $D$
\begin{equation}\label{density}
D=\frac{1}{N!}\sum_{i=0}^{N!-1}X^i=\frac{1}{N}J_{N},
\end{equation}
where $J_{N}$ is $N\times N$ matrix with all matrix elements equal
to 1.

Due to the matrices $X^i, (i=0,1,2,3,...,N!-1)$ have unique one in
every row and column, one can prove that $D$ commutes with $X^i,
(i=0,1,2,...,N!-1)$, i.e.,
\begin{equation}
[D, X^i]=0 \quad \mbox{for}\ (i=0,1,2,...,N!-1).
\end{equation}
Therefore we have
\begin{equation}
D=(1-\sum_{i=1}^{N!-1}p_i)X^0DX^{0\dagger}+\sum_{i=1}^{N!-1}p_iX^iDX^{i\dagger},
\end{equation}
which is independent on parameter $p_i$ and $X^0$ is identity matrix
or not flipping operator.

Now to obtain the unitary transformation of Bob  strategy we need
eigenvector of density matrix $D$. The density matrix $D$ is
circulant matrix \cite{Zhang}. An important property of circulant
matrices is that they (unitarily) diagonalizable by the Fourier
matrix
\begin{equation}\label{Fourier}
F=\frac{1}{\sqrt{n}}V(\omega)
\end{equation}
where $\omega=e^{2\pi i/n}$ and $V(\omega)$ is the Vandermonde
matrix defined as
\begin{equation}
V(\omega)=\left(
  \begin{array}{ccccc}
    1 & 1 & 1 & \cdots & 1 \\
    1 & \omega & \omega^2 & \cdots & \omega^{n-1} \\
    1 & \omega^2 & \omega^4 & \cdots & \omega^{2(n-1)} \\
    \vdots & \vdots &\vdots & \vdots & \vdots \\
    1 & \omega^{n-1} & \omega^{2(n-1)} & \cdots  & \omega^{(n-1)^2} \\
  \end{array}
\right).
\end{equation}
Let the $k$-th column vector of $V(\omega)$ be denoted by
$\ket{\omega_k}$, $k=0,1,2,...,n-1$. It is easy to verify that $F$
is unitary (i.e., $F^{-1}=F^{\dagger}$), since the Vandermonde
matrix obeys $V(\omega)^{-1}=V(\omega^{-1})$. Thus, for a $n$-square
circulant matrix $C$ we have
\begin{equation}\label{e2}
F^{\dagger}CF=diag(f(\omega^0), f(\omega^1),...,f(\omega^{(n-1)}))
\end{equation}
where $f(\mu)=c_0+c_1\mu^1+c_1\mu^2+...+c_1\mu^{(n-1)}$ and
$(c_0,c_1,...,c_{n-1})$ are matrix elements of $C$.

 In this case, the  matrix elements are fixed and equal to
 $\frac{1}{N}$, i.e., $c_0=c_1=...=c_{N-1}=\frac{1}{N}$, where
 the quantity of $n$ is $N$. Therefore, by using Eq.(\ref{e2}) and
 above definition for $f$, one can obtain eigenvalues of density matrix of $D$
 \begin{equation}
 \lambda_0=f(\omega^0)=1, \quad  \lambda_1=f(\omega^1)=
 \lambda_2=f(\omega^2)=\cdots=
 \lambda_{N-1}=f(\omega^{N-1})=0,
 \end{equation}
with the eigenvectors as
\begin{equation}
\ket{\lambda_k}=\frac{1}{\sqrt{N}}\ket{\omega_k}, \quad \mbox{for} \
k=0,1,2,...,N-1,
\end{equation}
where $\ket{\omega_k}$ denote the $k$-th column vector of
$V(\omega)$.

 In the above calculation we used formula:
\begin{equation}
\sum_{l=0}^{N-1}\omega^{lk}=0, \quad \mbox{for} \ \omega=
e^{\frac{2\pi i}{N-1}}.
\end{equation}
Now we are in the position to derive the unitary transformation of
Bob strategy as
$$
T_1=(\ket{\lambda_0}, \ket{\lambda_1},..., \ket{\lambda_{N-1}}
)=\frac{1}{\sqrt{N}}V(\omega)
$$
$$
T_2=(\ket{\lambda_{N-1}}, \ket{\lambda_0},...,
\ket{\lambda_{N-2}})=\frac{1}{\sqrt{N}}\left(
  \begin{array}{ccccc}
    1 & 1 & 1 & \cdots & 1 \\
    \omega^{N-1} & 1 & \omega & \cdots & \omega^{N-2} \\
    \omega^{2(N-1)} & 1 & \omega^2 & \cdots & \omega^{2(N-2)} \\
    \vdots & \vdots &\vdots & \vdots & \vdots \\
    \omega^{(N-1)^2} & 1 & \omega^{(N-1)} & \cdots  & \omega^{(N-1)(N-2)} \\
  \end{array}
\right)
$$
$$
\qquad \qquad\qquad\qquad{\vdots}
$$$$
\qquad \qquad\qquad\qquad\vdots
$$
\begin{equation}\label{Trans}
T_{N}=(\ket{\lambda_1}, \ket{\lambda_2},...,
\ket{\lambda_{N-1}},\ket{\lambda_{0}})=\frac{1}{\sqrt{N}}\left(
  \begin{array}{ccccc}
    1 & 1 & \cdots & 1 & 1 \\
    \omega & \omega^2  &\cdots & \omega^{N-1} &1 \\
    \omega^{2} & \omega^4 & \cdots & \omega^{2(N-1)} & 1 \\
    \vdots & \vdots &\vdots & \vdots & \vdots \\
    \omega^{(N-1)} & \omega^{2(N-1)} & \cdots & \omega^{(N-1)^2}  & 1 \\
  \end{array}
\right).
\end{equation}
Now we consider quantum roulette game with arbitrary $N$ states and
we discuss Bob and Alice how to control the game by quantum and
classical strategies, respectively. We can consider quantum roulette
as Hilbert space with $N$-dimension. The  initial states which can
adopt are
$$
\ket{1}\bra{1}=\left(
  \begin{array}{ccccc}
    1 & 0 & \cdots & 0 & 0 \\
    0& 0  &\cdots & 0 & 0 \\
    0 & 0 & \cdots & 0 & 0 \\
    \vdots & \vdots &\vdots & \vdots & \vdots \\
    0 & 0 & \cdots & 0  & 0 \\
  \end{array}
\right), \quad  \ket{2}\bra{2}=\left(
  \begin{array}{ccccc}
    0 & 0 & \cdots & 0 & 0 \\
    0& 1  &\cdots & 0 & 0 \\
    0 & 0 & \cdots & 0 & 0 \\
    \vdots & \vdots &\vdots & \vdots & \vdots \\
    0 & 0 & \cdots & 0  & 0 \\
  \end{array}
\right),
$$
$$
\qquad \qquad\qquad\qquad{\vdots}
$$$$
\qquad \qquad\qquad\qquad\vdots
$$
\begin{equation}
\ket{N}\bra{N}=\left(
  \begin{array}{ccccc}
    0 & 0 & \cdots & 0 & 0 \\
    0& 0  &\cdots & 0 & 0 \\
    0 & 0 & \cdots & 0 & 0 \\
    \vdots & \vdots &\vdots & \vdots & \vdots \\
    0 & 0 & \cdots & 0  & 1 \\
  \end{array}
\right),
\end{equation}
where one can obtain  above density matrices form similarity
transformation i.e., $\ket{k}\bra{k}=T_k^{\dagger}DT_k$ for
$k=1,2,...,N$.

We suppose that the initial state of the roulette is $\ket{\psi_0}$
and its density matrix is $\rho_0=\ket{\psi_0}\bra{\psi_0}$. Then
Alice and Bob will play a roulette game. During the game, Bob adopts
a quantum strategy by using a unitary matrix to act on the coins
while Alice  adopts the usual classical probabilistic
strategy. Now we have:\\
\textbf{Step1:} \ Alice places the roulette on one box such that the
state of the coins is known by both Alice and Bob. \\
\textbf{Step2:}\ Bob uses a unitary transformation $U_1$  to act on
the roulette, such that, if the initial state is
$\ket{1},\ket{2},...,\ \mbox{or} \ \ket{N}$ the unitary
transformation $U_1$ is corresponding $T_1, T_2,..., \ \mbox{or} \
T_{N}$, then the state of the coins become
\begin{equation}
\rho_1=U_1 \rho_0 U_1^{\dagger}=D
\end{equation}
\textbf{Step3:} Alice continues to play by employing classical
strategy, namely she perhaps changes the state of the roulette using
the permutation matrices $X^i \ (i=0,1,2,..., N!-1)$ with the
probability $p_i$. Thus, at the end of Alice's play, the state of
the roulette is described by the density matrix
\begin{equation}\label{alice}
\rho_2=(1-\sum_{i=1}^{N!-1}p_i)X^0\rho_1X^{0\dagger}+
\sum_{i=1}^{N!-1}p_iX^i\rho_1X^{i\dagger}=(1-\sum_{i=1}^{N!-1}p_i)X^0DX^{0\dagger}+
\sum_{i=1}^{N!-1}p_iX^iDX^{i\dagger}=D.
\end{equation}
\textbf{Step4:} Bob plays with the roulette by using the unitary
transformation $U_2$ so that the density matrix of final state of
the coins is given by
\begin{equation}
\rho_3=U_2 \rho_2 U_2^{\dagger}.
\end{equation}
Here Bob can get every arbitrary state with choice
$U_2=U_1^{\dagger}$, for example if he wants to get the state
$\ket{1}$ then $U_2=T_1^{\dagger}$; and $\ket{1}$,
$U_2=T_2^{\dagger}$ and so forth. Therefore, Bob can control the
game by the above quantum strategies.

This strategy is general method and one can employ easily to any
integer $N$. For example we employ it for quantum roulette with
$N=3$. In this case, there is a choice  of $3!$ possible flips
corresponding to all group symmetric $S_3$. Thus the basis and
permutation matrices \cite{sagan} are
$$
\ket{1}=\left(
          \begin{array}{c}
            1 \\
            0 \\
            0 \\
          \end{array}
        \right), \quad \ket{2}=\left(
          \begin{array}{c}
            0 \\
            1 \\
            0 \\
          \end{array}
        \right), \quad \ket{3}=\left(
          \begin{array}{c}
            0 \\
            0 \\
            1 \\
          \end{array}
        \right),
$$
$$
X^0=\left(
      \begin{array}{ccc}
        1 & 0 & 0 \\
        0 & 1 & 0 \\
        0 & 0 & 1 \\
      \end{array}
    \right), \quad X^1=\left(
      \begin{array}{ccc}
        0 & 1 & 0 \\
        1 & 0 & 0 \\
        0 & 0 & 1 \\
      \end{array}
    \right), \quad X^2=\left(
      \begin{array}{ccc}
        0 & 0 & 1 \\
        0 & 1 & 0 \\
        1 & 0 & 0 \\
      \end{array}
    \right), \quad
$$
\begin{equation}
X^3=\left(
      \begin{array}{ccc}
        1 & 0 & 0 \\
        0 & 0 & 1 \\
        0 & 1 & 0\\
      \end{array}
    \right), \quad X^4=\left(
      \begin{array}{ccc}
        0 & 0 & 1 \\
        1 & 0 & 0 \\
        0 & 1 & 0 \\
      \end{array}
    \right), \quad X^5=\left(
      \begin{array}{ccc}
        0 & 1 & 0 \\
        0 & 0 & 1 \\
        1 & 0 & 0 \\
      \end{array}
    \right),
\end{equation}
respectively. Then by using the Eq. (\ref{density}), we can
construct the density matrix $D$ as
\begin{equation}
D=\frac{1}{3!}\sum_{i=0}^{5}X^i=\frac{1}{3}J_3=\frac{1}{3}\left(
                                                            \begin{array}{ccc}
                                                              1 & 1 & 1 \\
                                                              1 & 1 & 1 \\
                                                              1 & 1 & 1 \\
                                                            \end{array}
                                                          \right).
\end{equation}
The Fourier matrix (\ref{Fourier}) and unitary transformation $T_i$
($i=1,2,3$)Eq. (\ref{Trans}) are given by
$$
F=\frac{1}{\sqrt{3}}\left(
  \begin{array}{ccc}
    1 & 1 & 1  \\
    1 & \omega & \omega^2  \\
    1 & \omega^2 & \omega^4  \\
      \end{array}
\right).
$$
\begin{equation}
T_1=F, \quad T_2=\frac{1}{\sqrt{3}}\left(
  \begin{array}{ccc}
    1 & 1 & 1  \\
    \omega^2 & 1 & \omega  \\
    \omega^4 & 1 & \omega^2  \\
      \end{array}
\right), \quad T_3=\frac{1}{\sqrt{3}}\left(
  \begin{array}{ccc}
    1 & 1 & 1  \\
    \omega & \omega^2 & 1  \\
    \omega^2 & \omega^4 & 1  \\
      \end{array}
\right),
\end{equation}
for $\omega=e^{\frac{2\pi i}{3}}$.

For initial state $\ket{\psi_0}$, i.e., $\ket{1}, \ket{2}$ or
$\ket{3}$, the unitary transformation $U_1$  is corresponding $T_1,
T_2,$ or $T_3$. For example we consider initial state
$\ket{\psi_0}=\ket{2}$ then the density matrix $\rho_0$ is
\begin{equation}
\rho_0=\ket{2}\bra{2}=\left(
  \begin{array}{ccc}
    0 & 0 & 0  \\
    0 & 1 & 0  \\
    0 &0 & 0  \\
      \end{array}
\right).
\end{equation}
Thus Bob rotates the initial density matrix as
\begin{equation}
\rho_1=U_1\rho_0U_1^{\dagger}=T_2\rho_0T_2^{\dagger}=D.
\end{equation}
Alice continues to play by employing classical strategy, then by
using Eq. (\ref{alice}) the density matrix takes the form
$$
\rho_2=(1-(p_1+p_2+p_3+p_4+p_5))X^0\rho_1X^{0\dagger}+p_1X^1\rho_1X^{1\dagger}+p_2X^2\rho_1X^{2\dagger}
$$
\begin{equation}
+p_3X^3\rho_1X^{3\dagger}+p_4X^4\rho_1X^{4\dagger}+p_5X^5\rho_1X^{5\dagger}=D.
\end{equation}
Bob can always control the final state by using the unitary matrix
$U_2=T_1^\dagger,T_2^\dagger,T_3^\dagger $, i.e.,\\
if he want state $\ket{1}$
\begin{equation}
\rho_3=T_1^\dagger\rho_2T_1=\left(
                              \begin{array}{ccc}
                                1 & 0 & 0 \\
                                0 & 0 & 0 \\
                                0 & 0 & 0 \\
                              \end{array}
                            \right)=\ket{1}\bra{1},
\end{equation}
if he want state $\ket{2}$
\begin{equation}
\rho_3=T_2^\dagger\rho_2T_2=\left(
                              \begin{array}{ccc}
                                0 & 0 & 0 \\
                                0 & 1 & 0 \\
                                0 & 0 & 0 \\
                              \end{array}
                            \right)=\ket{2}\bra{2},
\end{equation}
if he want state  $\ket{3}$
\begin{equation}
\rho_3=T_3^\dagger\rho_2T_3=\left(
                              \begin{array}{ccc}
                                0 & 0 & 0 \\
                                0 & 0 & 0 \\
                                0 & 0 & 1 \\
                              \end{array}
                            \right)=\ket{3}\bra{3}.
\end{equation}

\subsection{Noisy quantum game}
In this subsection we investigate the interaction between the
quantum game and its environment. The interaction between the system
and its environment introduce decoherence to the system, which is a
process of the undesired correlation between the system and the
environment when the system evolves. Therefore, the communication
accomplished under noisy channels \cite{Schumacher} may not be
faithful because the receiver may obtain partial or corrupted
information different from sender's information. The quantum noise
process is represented by mapping $\rho \Longrightarrow
\mathcal{S}(\rho)$, where $\mathcal{S}$ is a super-operator
\cite{Kraus} that makes the initial state $\rho$ evolve to the final
state $\mathcal{S}(\rho)$. In general, the communication process of
an open system can be represented by the operator-sum representation
\begin{equation}
{\mathcal{S}(\rho)}=\sum_{k} E_k\rho E_k^{\dagger},
\end{equation}
where $E_k$ are kraus operator elements for the super operation
$\mathcal{S}$, and are trace-preserving, $\sum_{k} E_k
E_k^{\dagger}=I$. There are several decoherence channels which for
example we consider the depolarizing channel. For a qubit system the
kraus operators of depolarizing channel are
\begin{equation}
E_0=\sqrt{1-r} I,\quad E_1=\sqrt{\frac{r}{3}} \sigma_x, \quad
E_2=\sqrt{\frac{r}{3}} \sigma_y, \quad E_3=\sqrt{\frac{r}{3}}
\sigma_z
\end{equation}
where $(\sigma_x,\sigma_y,\sigma_z)$ is the set  of Pauli matrices.
This channel acts on qubits by phase flips, amplitude flips or
combinations of both applied with probability $r/3$ each. More
generally, for a quantum $d$-ary digit (a qudit), is a
$d$-dimensional Hilbert space $H$ with orthonormal basis as
$(\ket{0}, \ket{1},...,\ket{d-1})$ , we can define a depolarizing
channel \cite{Barg} as follows:
\begin{equation}\label{depo}
E_0=\sqrt{1-r} I,\quad E_k=\sqrt{\frac{r}{d^2-1}}M_{i,j}, \quad i,
j\in {\mathcal{F}}_d
\end{equation}
where ${\mathcal{F}}_d$ is a finite field and $M_{i,j}$ is defined
as $\{M_{i,j}=Y^iZ^j, i,j\in {\mathcal{F}}_d\}$ such that
\begin{equation}\label{noise}
Y\ket{l}=\ket{(l-1) \mbox{mod d}}, \quad  Z\ket{l}=\omega^l\ket{l}
\end{equation}
and $\omega=e^{\frac{2 \pi i}{d}}$ is a primitive $d$th root of
unity.

Now we consider the  previous  example (i.e., $N=3$) with
decoherence model. In this case  the depolarizing channel has $E_k$
(Eq. \ref{depo}) represented by
$$
E_0=\sqrt{1-r} I, \quad E_1=\sqrt{\frac{r}{8}}Y, \quad
E_2=\sqrt{\frac{r}{8}}Z, \quad E_3=\sqrt{\frac{r}{8}}Y^2,\quad
E_4=\sqrt{\frac{r}{8}}YZ,
$$
\begin{equation}
E_5=\sqrt{\frac{r}{8}}Y^2Z,\quad E_6=\sqrt{\frac{r}{8}}YZ^2,\quad
E_7=\sqrt{\frac{r}{8}}Y^2Z^2,\quad E_8=\sqrt{\frac{r}{8}}Z^2,
\end{equation}
with
\begin{equation}
Y=\left(
  \begin{array}{ccc}
    0 & 1 & 0 \\
    0 & 0 & 1 \\
    1 & 0 & 0 \\
  \end{array}
\right), \quad\quad Z=\left(
                        \begin{array}{ccc}
                          1 & 0 & 0 \\
                          0 & \omega & 0 \\
                          0 & 0 &  \omega^2\\
                        \end{array}
                      \right), \quad \omega=e^{\frac{2 \pi i}{3}}.
\end{equation}
For simplicity, we consider only the interaction between the system
with its environment in the first step. Therefore, we have

 \textbf{Step1.}  The initial state $\rho_0=\ket{2}\bra{2}$ is decohered by
 depolarizing channel where become
 \begin{equation}
\acute{\rho_0}=\sum_{k=0}^{8}E_k\rho_0
E_k^{\dagger}=\frac{r}{8}\left(
\begin{array}{ccc}
1 & \alpha+e^{\frac{-2\pi i}{3}} & 1 \\
\alpha+e^{\frac{2\pi i}{3}}& \frac{8}{r}-7-\alpha & \alpha+e^{\frac{2\pi i}{3}} \\
1 & \alpha+e^{\frac{-2\pi i}{3}} & 1 \\
\end{array}
\right)
\end{equation}
where $\alpha=2\sqrt{2r(1-r)}$.

\textbf{Step2.} Bob the applies the unitary transformation $U_1=T_2$
as
\begin{equation}
\rho_1=T_2\acute{\rho_0}T_2^\dagger=\left(
                                      \begin{array}{ccc}
                                        \beta & \xi &\xi \\
                                        \xi^\star & \eta & \eta \\
                                       \xi^\star & \eta & \eta \\
                                      \end{array}
                                    \right),
\end{equation}
with
$$
\beta=1/24 (8 + (-5 + 3 \alpha) r), \quad \xi=1/48 (16 + (-19 - 3
i\sqrt{3}) r), \quad \eta=1/24 (8 - (5 + 3 \alpha) r).
$$
In this case we see that  the $\rho_1$ is not equal with $D$.

\textbf{Step3.} Now the Alice does not delay (i.e., the system is
not decohered) and continues to play by employing classically
strategy, then we have
\begin{equation}
\rho_2=(1-\sum_{i=1}^5p_i)X^0\rho_1X^{0\dagger}+\sum_{i=1}^5p_iX^i\rho_1X^{i\dagger}\neq
D
\end{equation}
\textbf{Step4.} Then Bob implements the unitary transformation
$U_2=T_2^\dagger$. In this case the density matrix of final state is
\begin{equation}
\rho_3=T_2^\dagger \rho_2T_2=\left(
                               \begin{array}{ccc}
                                 \rho_{11} & \rho_{12} & \rho_{13} \\
                                 \rho_{21} & \rho_{22} & \rho_{23} \\
                                 \rho_{31} & \rho_{32} & \rho_{33} \\
                               \end{array}
                             \right)
\end{equation}
where
$$
\rho_{11}=-\frac{r}{16} (-2 +
   2 p_1 + (3 +i \sqrt{3}) p_4 +
   3 p_5 - i \sqrt{3} p_5)
  $$
$$
\rho_{12}=\frac{re^{2 i\pi/3}}{16} (-2 +
   \alpha - i\sqrt{3}
     \alpha + (5 - i\sqrt{3} + (-1 + 3 i\sqrt{3}) \alpha)p_1 + (3 + i\sqrt{3} -
      3 \alpha + i\sqrt{3} \alpha) p_2 +
$$$$
   (3  - i\sqrt{3}  +
   2i\sqrt{3} \alpha )p_4 +
   (3  + i\sqrt{3}  -
   3 \alpha  + i\sqrt{3}\alpha) p_5)
$$
$$
\rho_{13}=\frac{re^{2 i\pi/3}}{16\sqrt{3}} (-3 i - \sqrt{
   3} + (9 i + \sqrt{
   3}) p_1 +
   3 (i + \sqrt{
   3}) p_2))
$$
$$
\rho_{21}=-\frac{r}{16 \sqrt{3}} (-3 i + \sqrt{3} -
    2 \sqrt{3}
      \alpha + (6 i -
       4 \sqrt{3} + (3 i + 5 \sqrt{3}) \alpha) p_1 +
    3 (2 i + (-i + \sqrt{3}) \alpha) p_2 +
    $$$$
    ((6  -3\alpha )i + 3 \sqrt{3} \alpha) p_4 +
    (3( 1+\alpha)i+ 3 \sqrt{3}(-1+\alpha) )p_5 )
$$
$$
\rho_{22}=\frac{1}{8} (-8 + (7 + \alpha) r) (-1 + p_1)
$$
$$
\rho_{23}=-\frac{r}{16 \sqrt{3}} (-3 i +\sqrt{3} -
    2 \sqrt{3}
\alpha + (9 i - \sqrt{3} + (-3 i + 5\sqrt{3})\alpha) p_1 +
    3 (i - \sqrt{3} + (i + \sqrt{3})\alpha) p_2 +
$$$$
    (3 (1+\alpha)i +3 \sqrt{3}(-1 +
     \alpha ))p_4 +
    (6 (3-\alpha)i  +
    3 \sqrt{3} \alpha )p_5))
$$
$$
\rho_{31}=\frac{r}{16 } (2 + (-5 + \sqrt{3}i) p_1 + (-3 - i
\sqrt{3}) p_2)
$$
$$
\rho_{32}=\frac{r}{16 } (-1 - i \sqrt{3} +
   2 \alpha + (1 +
      3 i\sqrt{3} + (-5 - i \sqrt{3})\alpha) p_1 + (3 + i\sqrt{3}- 3\alpha + i \sqrt{3} \alpha) p_2 +
      $$$$
   (\sqrt{3}(2 - \alpha )i -
   3 \alpha) p_4  +( \sqrt{3}(1+\alpha)i+3(1-\alpha))p_5 )
$$
$$
\rho_{33}=-\frac{r}{16 } (-2 +
   2 p_1 + (3 - i \sqrt{3})p_4 +
   (3 + i \sqrt{3}) p_5)
   $$
where it is easy to work out the probability of getting a state
$\ket{2}$ at the end of the game, and equal with $\frac{1}{8} (-8 +
(7 + \alpha) r) (-1 + p_1)$.

\section{Conclusion}
We have introduced the method to discuss the quantum game of
roulette with arbitrary $N$ state. In this method, we can get a
matrix $D$ which the density matrix of initial state of roulette can
be changed to after Bob using a proper unitary transformation $U_1$,
that is $\rho_1=D$. And then, Bob can use another proper unitary
transformation $U_2$ to control the game because this matrix $D$ is
invariant under the classical transposition. This method is general
and one can employ to any finite-dimensional quantum game. Finally,
we have considered  the interaction between the system and its
environment and investigate depolarizing channel on this strategy.
In this case, we have shown that if  Bob cannot control the noise in
the system completely, he stands to lose the advantage through the
utilization of quantum devices.

\end{document}